\documentclass[12pt]{elsarticle}

\usepackage{graphicx}
\usepackage{amssymb}
\usepackage{amsmath}
\usepackage{lineno}

\usepackage{overpic}
\usepackage{xcolor}
\usepackage{booktabs}
\usepackage{hyperref}

\journal{JQSRT. \textcopyright 2017. \href{http://creativecommons.org/licenses/by-nc-nd/4.0/}{CC-BY-NC-ND 4.0} }

\newcommand{\lmax}{l_\mathrm{max}}
\newcommand{\kmax}{\kappa_\mathrm{max}}
\newcommand{\scat}{\mathrm{scat}}
 
\newcommand{\ui}{\mathrm{i}} 
 
\newcommand{\fatE}{\mathbf{E}}
\newcommand{\fatr}{\mathbf{r}}
\newcommand{\ofr}{\left(\mathbf{r}\right)}
\newcommand{\fatM}{\mathbf{M}}
\newcommand{\fatd}{\hat{\mathbf{d}}}
\newcommand{\fatj}{\mathbf{j}}
\newcommand{\unite}{\hat{\mathbf{e}}}
\newcommand{\nm}{\,\mathrm{nm}}
\DeclareMathOperator{\MSE}{MSE}

\begin{document}

\begin{frontmatter}

\title{Extending the applicability of the T-matrix method to light scattering by flat particles on a substrate via truncation of Sommerfeld integrals}

\author[lti,imt]{Amos Egel}
\author[mos]{Yuri Eremin}
\author[iwt]{Thomas Wriedt}
\author[lti]{Dominik Theobald}
\author[lti,imt]{Uli Lemmer}
\author[lti,imt]{Guillaume Gomard}

\address[lti]{Light Technology Institute, Karlsruhe Institute of Technology, 76131~Karlsruhe, Germany}
\address[imt]{Institute of Microstructure Technology, Karlsruhe Institute of Technology, 76131~	Karlsruhe, Germany}
\address[mos]{Lomonosov Moscow State University, ul. Leninskiye Gory 1, 119991 Moscow, Russia}
\address[iwt]{Institut f\"ur Werkstofftechnik, Badgasteiner Str. 3, 28359 Bremen, Germany}

\begin{abstract}
The simulation of light scattering by particles on a substrate with the $T$-matrix method relies on the expansion of the scattered field in spherical waves, followed by a plane wave expansion to allow the evaluation of the reflection from the substrate. 
In practice, the plane wave expansion (i.e., the Sommerfeld integrals) needs to be truncated at a maximal in-plane wavenumber $\kmax$. An appropriate selection of $\kmax$ is essential: counter-intuitively, the overall accuracy can degrade significantly if the integrals are truncated with a \emph{too large} value.
In this paper, we propose an empirical formula for the selection of $\kmax$ and discuss its application using a number of example simulations with dielectric and metallic oblate spheroids on dielectric and metallic substrates. The computed differential scattering cross sections are compared to results obtained from the discrete-sources method.
\end{abstract}


\end{frontmatter}

\section{Introduction}
\begin{figure}[t]
	\centering
	\begin{overpic}[width=70mm]
	{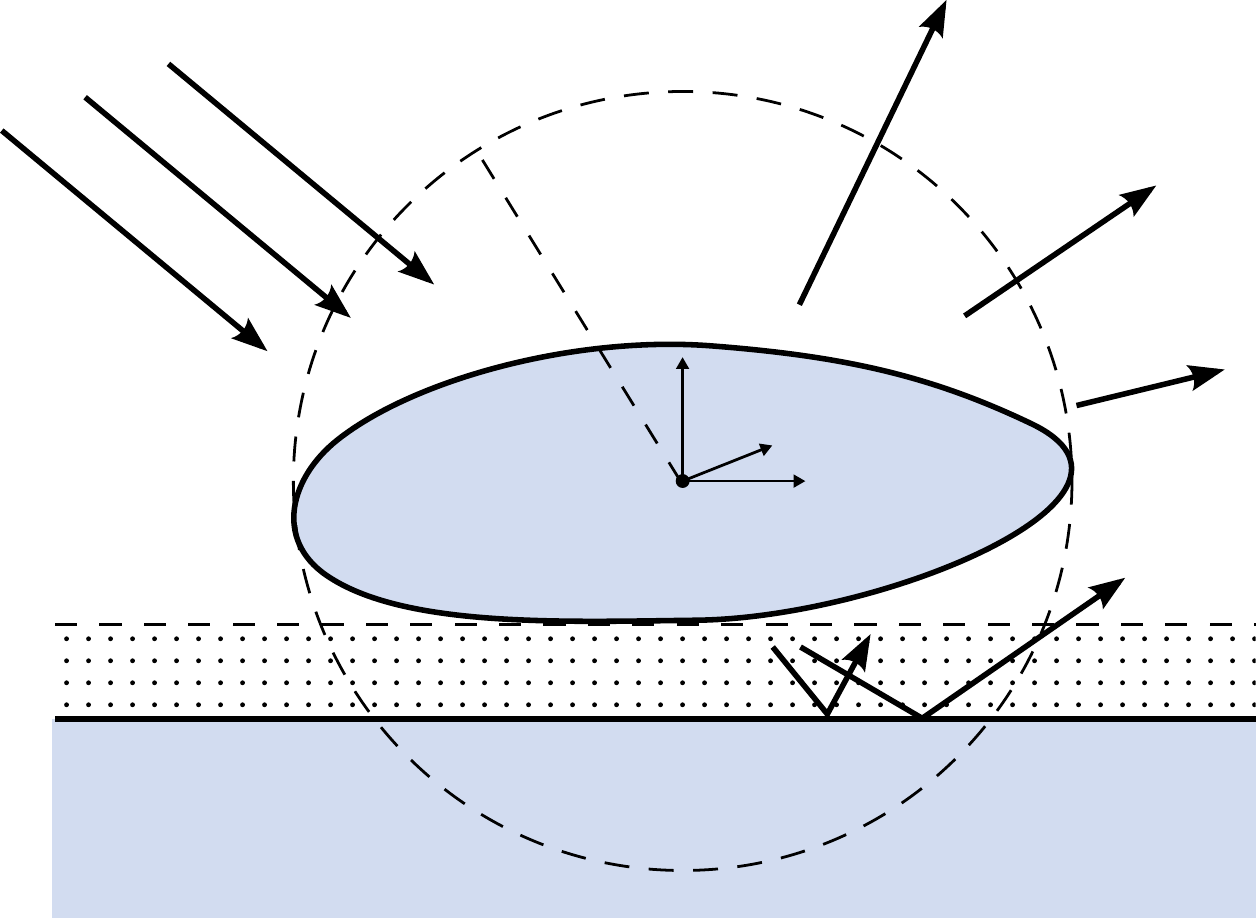}
    \put(10,25){\footnotesize $z_\mathrm{min}$}
    \put(43,55){\footnotesize $R$}
	\put(55,42){\footnotesize $z$}
	\put(65,35){\footnotesize $x$}
	\put(63,39){\footnotesize $y$}	
	\put(102,22){\footnotesize expand in}
	\put(102,16){\footnotesize plane waves}
	\put(20,68){\footnotesize initial field}
	\put(78,68){\footnotesize scattered field}	
	\end{overpic}	
	\caption{\label{fig:configuration}Scattering configuration. The dashed circle denotes the circumscribing sphere with radius $R$, and the horizontal dashed line denotes the bounding plane at $z=z_\mathrm{min}$. Below the particle (dotted region), the scattered field is expanded in downgoing plane waves.}
\end{figure}
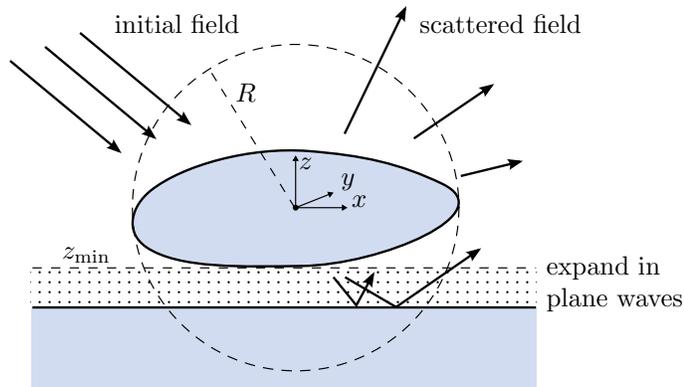

Light scattering by structures on a substrate is relevant in a variety of applications, including total internal reflection microscopy \cite{Helden2006}, surface enhanced Raman spectroscopy \cite{Ding2016} and quality control of silicon wafers \cite{Eremin1999}.
In optical simulations of these systems, it is important to take into account the particle-substrate scattering interaction. The scattered field from the particle is partially reflected by the substrate and is then incident on the particle again, compare figure \ref{fig:configuration}. Therefore, the optical response of the particle and the substrate cannot be regarded independently.

The $T$-matrix method introduced by Waterman \cite{Waterman_PotI_1965} is one of the most popular numerical techniques for the simulation of scattering by compact obstacles, and has been extended by Kristensson \cite{Kristensson1980} to the case of particles near infinite interfaces. 
In general, the scattered field is expanded in outgoing spherical vector waves originating from the particle center. With regard to the substrate reflection, these spherical waves need to be transformed into a plane wave expansion, allowing the application of Fresnel reflection for each partial plane wave. 

This transformation of the electric field's spherical wave expansion into a plane wave expansion typically enters the method in the form of one-dimensional integrals over the in-plane wavenumber $0\leq\kappa<\infty$. These so called Sommerfeld integrals are usually solved numerically, and the maximal wavenumber $\kmax$ at which the integrals are truncated is the central issue of this study. 

\begin{equation*}
\int_0^\infty \hspace{-2mm} \longrightarrow \int_0^{\kmax} \hspace{-5mm} 
\end{equation*}

In this paper, we focus on the critical case of flat particles on a substrate, when the particle's circumscribing sphere intersects with the planar surface. 
Due to the divergence of the spherical wave expansion inside the circumscribing sphere \cite{Auguie2016}, the applicability of the $T$-matrix method is not obvious for such geometries. 
Doicu et al. \cite{Doicu1999} have demonstrated numerically that in fact, a surprisingly good accuracy is possible,
and in a recent paper \cite{Egel2016}, we have argued that it is the convergence of the plane wave expansion and not of the spherical wave expansion that ensures a valid treatment of the reflected field from the interface. Thereby, it could be confirmed that the $T$-matrix method is in general valid also for flat particles on a substrate.  However, the spectrum of the plane wave expansion converges not uniformly, but only pointwise with increasing multipole order of the scattered field's original $T$-matrix representation \cite{Egel2016}. This has direct implications for the design of accurate numerical implementations.

Intuitively, one might expect that the largest $\kmax$ should lead to the best accuracy, and the selection of $\kmax$ was merely a trade-off between numerical effort and accuracy. 
But this is not the case: The overall numerical accuracy for fixed multipole order first improves with growing $\kmax$ and then drops rapidly -- a behavior that is also referred to as \emph{relative convergence} \cite{Mittra1972}. A sophisticated selection of $\kmax$ is thus essential to ensure the accuracy of the method \cite{Egel2016}.

Up to now, an a priori estimate for a suitable truncation wavenumber has been lacking. With this paper, we propose an empirical formula to estimate an appropriate $\kmax$ value as a function of the scattering particle size and of the truncation order $\lmax$ used for the particle $T$-matrix.

In order to obtain such an empirical formula, we study the convergence of the plane wave spectrum of a translated dipole source representing the induced currents in the outermost infinitesimal volume of the particle.
A similar approach has previously been applied  by Cappellin et al. \cite{Cappellin_RS_2008} in the context of antenna analytics.
We finally illustrate the method with the example of light scattering by an oblate spheroid on a substrate. The simulated differential scattering cross sections are compared to accurate baseline results computed with the discrete sources method (DSM) \cite{Eremin2007} to evaluate the accuracy of the results obtained using the proposed formula.

\begin{figure}[t]
	\centering
	\begin{overpic}[width=35mm]
		{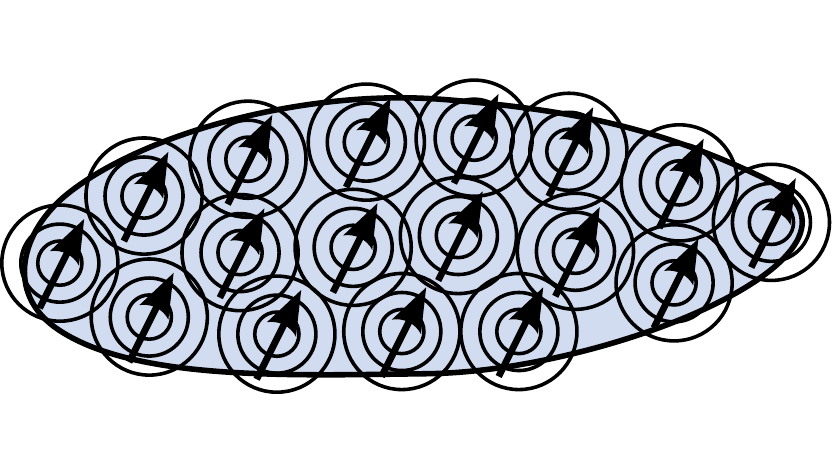}
		\put(10,65){physical picture}
	\end{overpic}	
	\hspace{5mm}
	\begin{overpic}[width=35mm]
		{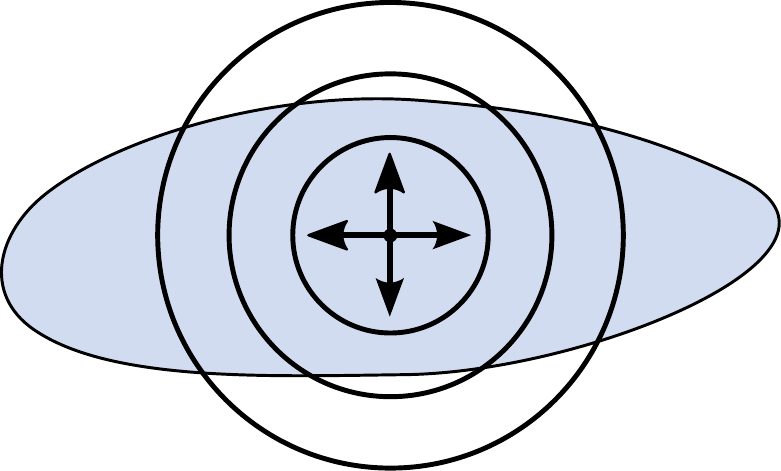}
		\put(10,65){$T$-matrix picture}	
	\end{overpic}
	\caption{\label{fig:vie vs tmatrix}Left: Physical picture according to the volume integral equation. Microscopic current sources are distributed over the scatterer. Right: $T$-matrix picture. The contributions of all microscopic currents are gathered to a multipole source in the particle center.}
\end{figure}

\section{The scattered field}
\label{sec:scattered field}
The scattered field of a particle in frequency domain can be thought of as the joint radiation from infinitesimal induced current sources $\mathbf{j}(\fatr) \, \mathrm{d}^3\fatr$ distributed over the particle volume $V$ \cite{Novotny2006}, see the left part of figure \ref{fig:vie vs tmatrix}:
\begin{equation}
\label{eq:vie}
\fatE_\scat(\fatr) = \int_V \hspace{-1mm} \mathrm{d}^3\fatr' \,\mathbf{G}(\fatr,\fatr') \cdot \mathbf{j}(\fatr'),
\end{equation}
where $\mathbf{G}(\fatr,\fatr')$ denotes the dyadic Greens function of the electric field.

\subsection{Spherical wave expansion}
The fundamental feature  \cite{Mishchenko2004} of the $T$-matrix method is the expansion of the scattered field in terms of outgoing spherical vector wave functions (SVWFs). For a particle centered at the coordinate origin,
\begin{align}
\label{eq:swe}
\fatE_\scat\ofr &= \sum_{p=1}^2 \sum_{l=1}^\infty \sum_{m=-l}^l b_{lmp} \, \fatM_{lmp}^{(3)} \ofr &\text{for }\left|\fatr\right| > R,
\end{align}
where $\fatM_{lmp}^{(3)}$ are the outgoing SVWFs of degree $l$, order $m$ and polarization $p$ (1 for spherical TE and 2 for spherical TM), and  $b_{lmp}$ are the expansion coefficients of the scattered field, whereas $R$ is the radius of the circumscribing sphere of the particle, see figure \ref{fig:configuration}. We use the definition of the SVWFs as provided in \cite{DoicuA2006}, with the difference that we distinguish between electric- and magnetic-type SVWFs using the polarization index $p$ (compare \cite{Hansen1988}), instead of using separate symbols $\mathbf{M}_{lm}^{1,3}$ and $\mathbf{N}_{lm}^{1,3}$ as in \cite{DoicuA2006}.

The spherical wave expansion coefficients of the scattered field can be constructed from \eqref{eq:vie} using the expansion of the dyadic Green's function in outgoing SVWFs \cite{DoicuA2006,Kristensson2016}:
\begin{align}
\label{eq:swe of G}
\mathbf{G}(\fatr,\fatr') &= \frac{\ui k}{\pi} \sum_{p=1}^2 \sum_{l=1}^\infty \sum_{m=-l}^l
\fatM_{lmp }^{(3)} (\fatr) \otimes  \fatM_{l, -m, p }^{(1)} (\fatr') &\text{for }\left|\fatr\right| > \left|\fatr'\right|
\end{align}
where $\fatM_{lmp}^{(1)}$ denotes the regular SVWFs and $k$ is the wavenumber of the medium in which the particle is located. Inserting \eqref{eq:swe of G} into \eqref{eq:vie} yields
\begin{align}
\label{eq:b}
b_{lmp} = \frac{\ui k}{\pi} \int_V \hspace{-1mm} \mathrm{d}^3\fatr' \, \fatM_{l, -m, p}^{(1)} (\fatr') \cdot \mathbf{j}(\fatr').
\end{align}

Note the conceptual difference between \eqref{eq:vie} and \eqref{eq:swe}. The contributions of the distributed infinitesimal currents in \eqref{eq:vie} (that are dipole sources) are gathered in \eqref{eq:swe} to one virtual source of multipole waves. In other words, all dipole sources are \emph{translated} to the particle center -- see also figure \ref{fig:vie vs tmatrix}. It is reasonable to expect (and we will see later) that this translation is numerically more delicate for infinitesimal currents $\mathbf{j}(\fatr) \, \mathrm{d}^3 \fatr$ far away from the particle center compared to those close to it. 

\subsection{Plane wave expansion}
In order to account for reflection from the substrate, the scattered field below the particle is expanded in terms of downgoing plane vector wave functions (PVWFs)
\begin{equation}
\fatE^-_j(\alpha, \kappa, \fatr) = \exp(\ui \mathbf{k}^- \cdot \fatr) \, \unite_j
\end{equation}
where $\mathbf{k}^-$ is a downgoing propagation vector (as the substrate is located below the particle) and $\unite_j$ is the unit vector in the polar ($j$=1) or azimuthal ($j=2$) direction. Thus, the PVWFs are parameterized by the in-plane wavenumber $\kappa$ and the azimuthal propagation angle $\alpha$, 
such that $(\kappa, \alpha, -k_z)$ with $k_z = (k^2-\kappa^2)^{1/2}$ are the cylindrical coordinates of $\mathbf{k}^-$, 
as well as the polarization parameter $j$ ($1$ for TE and $2$ for TM). 
The expansion of the scattered field in downgoing PVWFs therefore reads:
 
\begin{align}
\label{eq:pwe}
\fatE_\scat\ofr &= \sum_{j=1}^2 \int_0^{2\pi} \hspace{-2mm} \mathrm{d}\alpha \int_0^\infty \hspace{-2mm} \mathrm{d}\kappa \, \kappa \, g^-_j(\alpha,\kappa) \, \fatE^-_j (\alpha, \kappa; \fatr) &\text{for }z<z_\mathrm{min},
\end{align}
where $z=z_\mathrm{min}$ defines the transverse tangent plane that bounds the particle from below, see figure \ref{fig:configuration}. 
Then, $g^-_j(\alpha,\kappa)$ is the downgoing plane wave spectrum of the scattered field below the particle.

To construct the plane wave spectrum, one can use the expansion of the dyadic Green's function in downgoing PVWFs \cite{Bostrom1991,DoicuA2006}:
\begin{align}
\label{eq:pwe of G}
\mathbf{G}(\fatr,\fatr') &= \frac{\ui}{8\pi^{2}} \sum_{j=1}^2 \int_0^{2\pi} \hspace{-2mm} \mathrm{d}\alpha \int_0^\infty  \hspace{-2mm} \mathrm{d}\kappa \, \frac{\kappa}{k_z}\,\fatE_{j}^{-} (\kappa,\alpha;\fatr) \otimes \fatE_{j}^{-} (\kappa,\alpha;-\fatr') & \text{for }z < z'.
\end{align}
Inserting \eqref{eq:pwe of G} into \eqref{eq:vie} yields the exact downgoing plane wave spectrum for the current distribution $\mathbf{j}(\fatr)$:
\begin{align}
\label{eq:g from G}
g^-_j(\alpha,\kappa) &= \frac{\ui}{8\pi^{2} k_z}  \int_V \hspace{-1mm} \mathrm{d}^3\fatr' \, \fatE_{j}^{-} (\kappa,\alpha;-\fatr') \cdot \mathbf{j}(\fatr')
\end{align}
In numerical methods where the induced current distribution $\mathbf{j}(\fatr)$ is actually solved for (like the volume integral equation method \cite{Markkanen2012} or the discrete dipole approximation \cite{Draine1994}), \eqref{eq:g from G} could directly be used to numerically evaluate the plane wave spectrum. But in the $T$-matrix method, the scattered field coefficients are directly computed without solving for  $\mathbf{j}(\fatr)$, such that the scattered field's plane wave spectrum needs to be inferred from the spherical wave expansion coefficients $b_{lmp}$. This is done by making use of the expansion of outgoing SVWFs in downgoing PVWFs:

\begin{align}
\label{eq:SVWF in PVWF}
\fatM^{(3)}_{lmp} (\fatr) &= \sum_{j=1}^2 \int_0^{2\pi} \hspace{-2mm} \mathrm{d}\alpha \int_0^\infty \hspace{-2mm} \mathrm{d} \kappa \, \kappa \, B^-_{lmpj}(\alpha, \kappa) \, \fatE_j^-(\alpha, \kappa;\fatr) &\text{for }z<0,
\end{align}
where $B^-_{lmpj}(\alpha, \kappa)$ is the corresponding spherical to plane wave transformation operator \cite{Bostrom1991,Egel2016}. Inserting \eqref{eq:SVWF in PVWF} into \eqref{eq:swe} and using \eqref{eq:b} yields the spectral amplitude:
\begin{equation}
\label{eq:g from swe}
\begin{aligned}
g^-_j(\alpha,\kappa) &= 
\sum_{p=1}^2 \sum_{l=1}^\infty \sum_{m=-l}^l B^-_{lmpj}(\alpha, \kappa) \,b_{lmp}
\\ 
&= \frac{\ui k}{\pi} \sum_{p=1}^2 \sum_{l=1}^\infty \sum_{m=-l}^l B^-_{lmpj}(\alpha, \kappa) 
\int_V \hspace{-1mm} \mathrm{d}^3\fatr' \, \fatM_{l, -m, p}^{(1)} (\fatr') \cdot \mathbf{j}(\fatr').
\end{aligned}
\end{equation}
When constructing the plane wave expansion from the spherical wave expansion, the order of summation and integration is interchanged. In fact, this is not necessarily an identity operation and explains why the domain of validity for \eqref{eq:pwe} can be different from that of \eqref{eq:swe}, although the coefficients of the former are constructed from the latter. For a more detailed discussion of this aspect, see \cite{Egel2016}.	

\begin{figure}[t]
	\centering
	\begin{overpic}[width=0.4\textwidth]
		{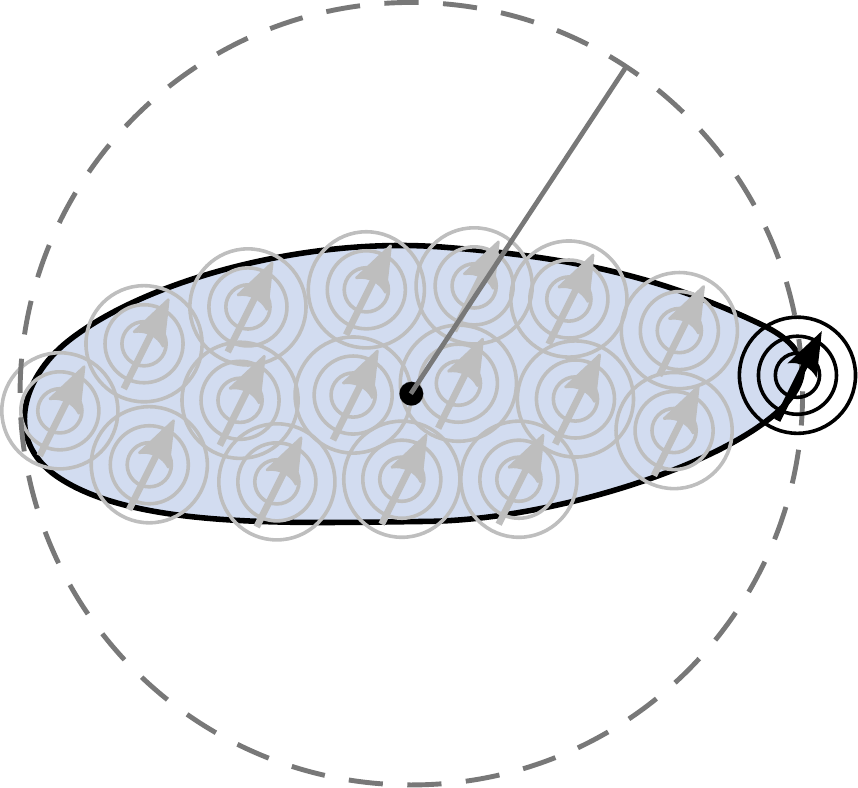}
        \put(57,54) {$R$}
	\end{overpic}
	\caption{\label{fig:single dipole}A single current element with maximal distance to the particle center, i.e., $|\fatr_d| = R$.}
\end{figure}

\section{Convergence of the plane wave spectrum}
\label{sec:convergence}

The reflection from the substrate can be directly computed from the plane wave expansion \eqref{eq:pwe} of the scattered field \cite{Egel2016}. It is thus the validity and accuracy of the scattered field's plane wave spectrum $g^-_j(\alpha,\kappa)$ that ensures the validity and accuracy of the reflected field from the substrate. For that reason, we will in the following section investigate the convergence of $g^-_j(\alpha,\kappa)$ without explicitly addressing the reflection from the substrate.

In section \ref{sec:scattered field}, we have cited two expressions for the plane wave spectrum of the scattered field of a particle: 
the exact expression \eqref{eq:g from G} and an expression \eqref{eq:g from swe} derived from the spherical wave expansion, which refers to the $T$-matrix method. 
The latter involves a series over the multipole order and degree, and here we want to study how the partial sums
\begin{equation}
\label{eq:g_lmax}
\begin{aligned}
g^-_{\lmax,j}(\alpha,\kappa) &= \frac{\ui k}{\pi}
\sum_{p=1}^2 \sum_{l=1}^{\lmax} \sum_{m=-l}^l B^-_{lmpj}(\alpha, \kappa) 
\int_V \hspace{-1mm} \mathrm{d}^3\fatr' \, \fatM_{l, -m, p}^{(1)} (\fatr') \cdot \mathbf{j}(\fatr').
\end{aligned}
\end{equation}
of this series converge to the exact plane wave spectrum with increasing truncation mutlipole order $\lmax$.

\begin{figure}[t!]
	\begin{overpic}[width=0.95\textwidth,tics=10]
		{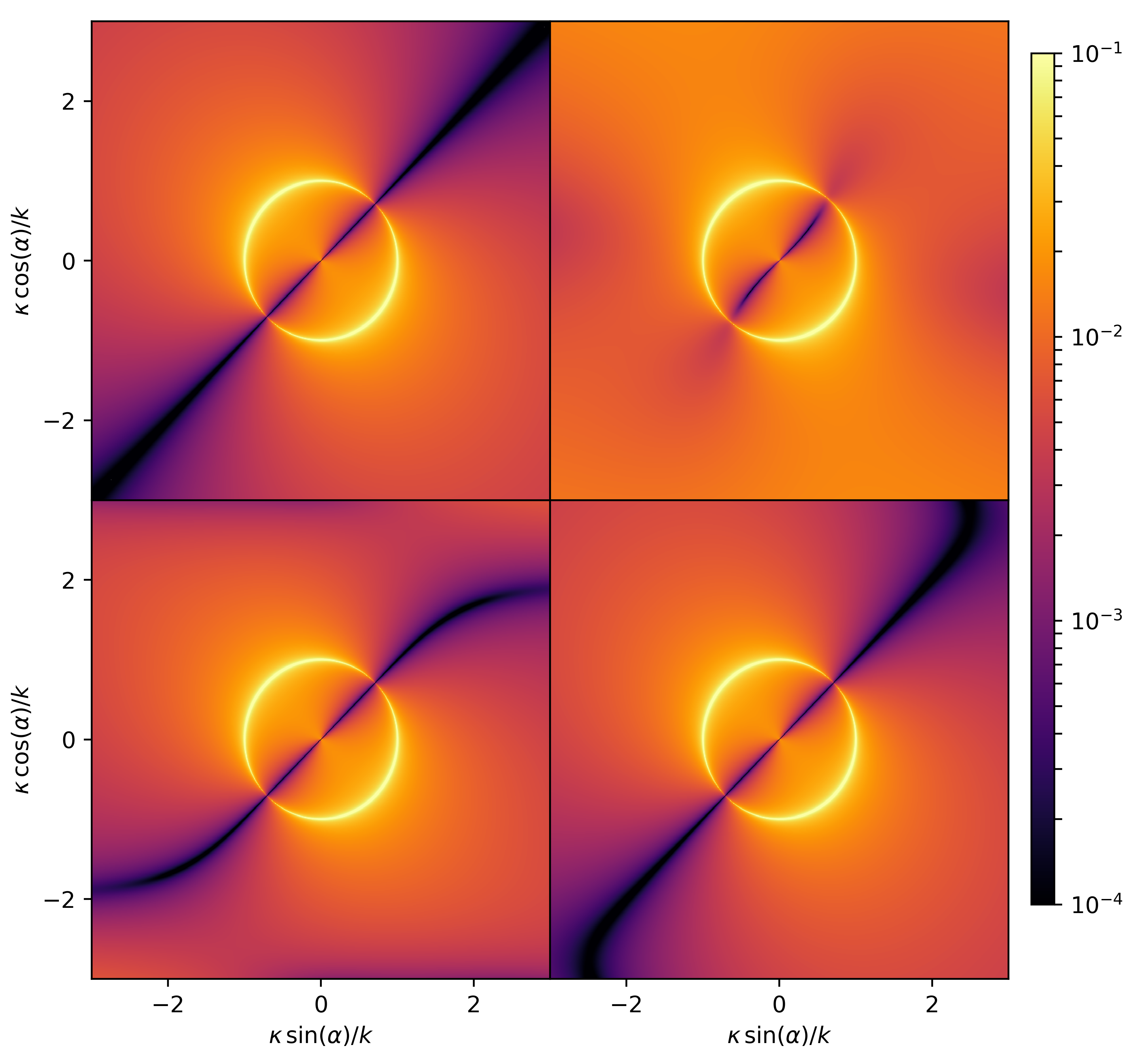}
		\put(10,87){exact}
		\put(50,87){$\lmax=2$}
		\put(10,45){$\lmax=5$}
		\put(50,45){$\lmax=8$}
	\end{overpic}
	\caption{\label{fig:g comparison}Modulus of the TE-polarized plane wave spectrum of a dipole source  with $k\fatr_d = (1, 0, 0)$ and $\fatd=(1,1,0)/\sqrt{2}$. Top left: exact spectrum. Other images: plane wave spectrum constructed from the spherical wave expansion with $\lmax=2$, $5$ and $8$.}
\end{figure}
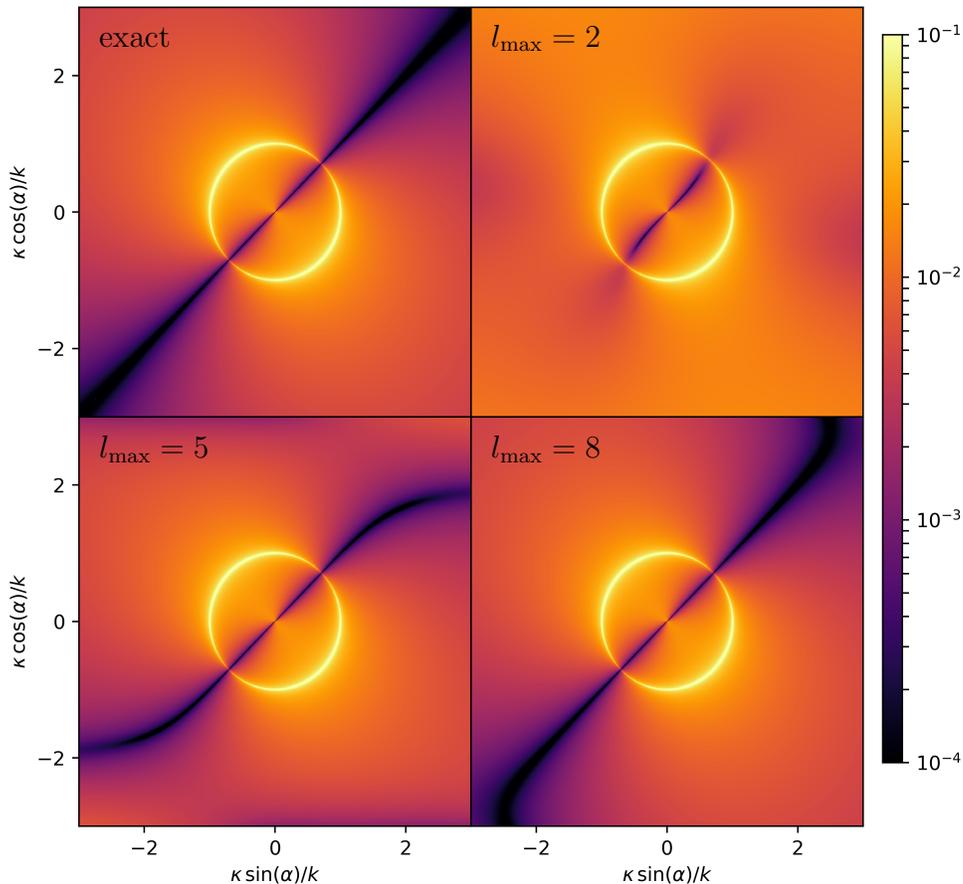

In order to quantify the error of a partial sum of \eqref{eq:g_lmax}, we define the azimuthally averaged relative quadratic error of the plane wave spectrum (MSE),
\begin{equation}
\MSE(\lmax, \kappa) = \frac{\sum_{j=1}^2\int_0^{2\pi} \mathrm{d} \alpha \left| g_{\lmax, j}^- (\alpha, \kappa) - g_j^-(\alpha, \kappa) \right|^2}{\sum_{j=1}^2\int_0^{2\pi} \mathrm{d} \alpha \left| g_j^-(\alpha, \kappa)\right|^2},
\end{equation}
\begin{figure}[h]
	\centering
	\begin{overpic}
	{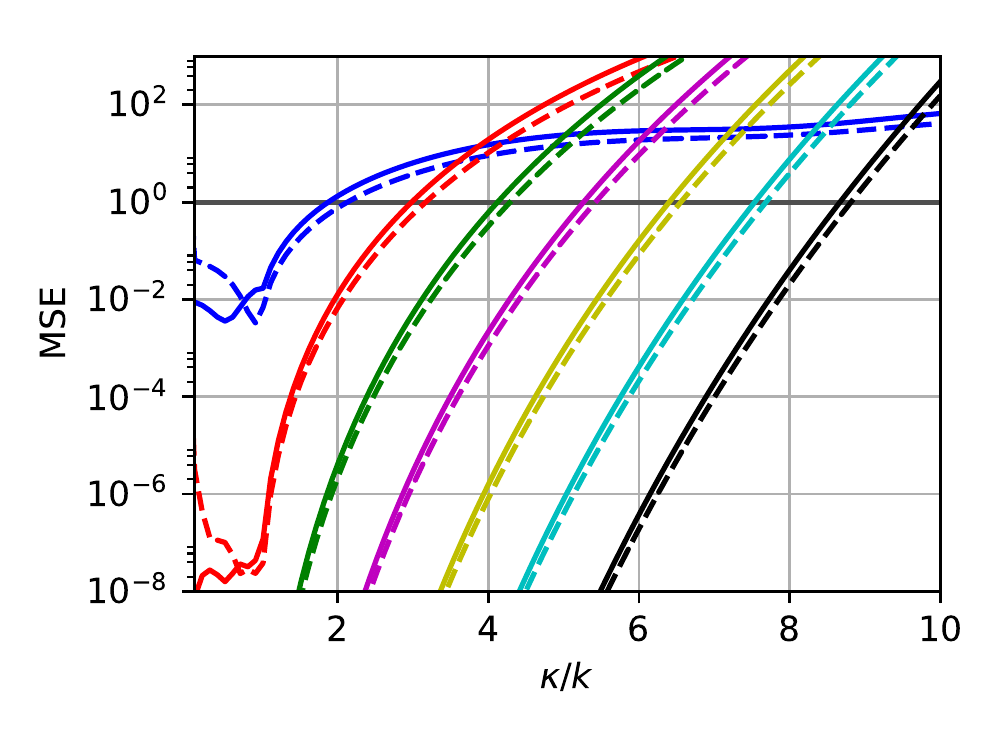}
	\put(25, 57){\footnotesize $\lmax=2$}
	\put(35, 49){\footnotesize $5$}
	\put(40, 44){\footnotesize $8$}
	\put(45, 40){\footnotesize $11$}
	\put(51, 37){\footnotesize $14$}
	\put(60, 37){\footnotesize $17$}
	\put(69, 37){\footnotesize $20$}			
	\end{overpic}
	\caption{Azimuthally averaged square error of plane wave spectrum for $k\fatr_d = (1, 0, 0)$ and $\fatd=(1,0,0)$ (solid lines) or $\fatd= (0,0,1)$ (dashed lines) for different values of $\lmax$.}
	\label{fig:quadratic residual}
\end{figure}
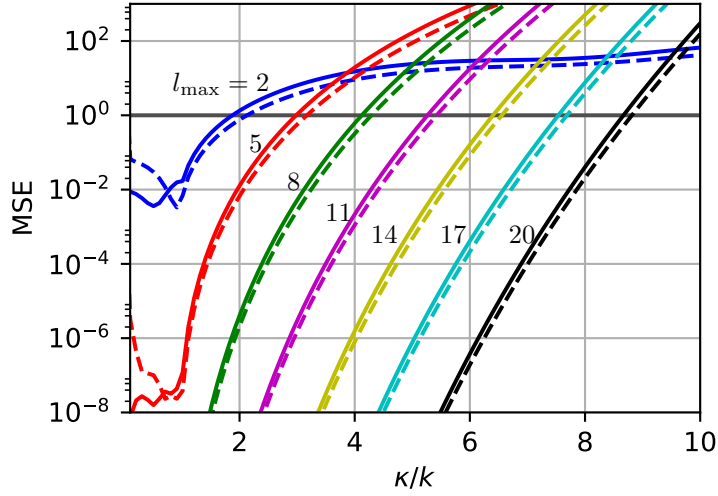

The basic idea is now to select the truncation scale $\kmax$ of the plane wave expansion such that $\MSE<1$ for all $\kappa<\kmax$:
\begin{equation}
\label{eq:kmax of lmax}
\kmax = \liminf \left\{ \kappa | \MSE(\lmax,\kappa)\geq 1 \right\}
\end{equation}
 Thus, we include only those $\kappa$-regions in the numerical treatment where the plane wave spectrum has already converged, whereas $\kappa$-regions where the convergence has not yet been achieved are excluded. In other words, we assume that when $\MSE>1$, it is numerically more favorable to just set the plane wave spectrum to zero as opposed to considering the erroneous contributions from the not-converged $g^-_{\lmax, j}(\alpha, \kappa)$.

\begin{figure}[t]
	\centering
	\graphicspath{{images/}}
	\def\svgwidth{0.7\textwidth}
	{\setlength{\fboxsep}{0.1pt} 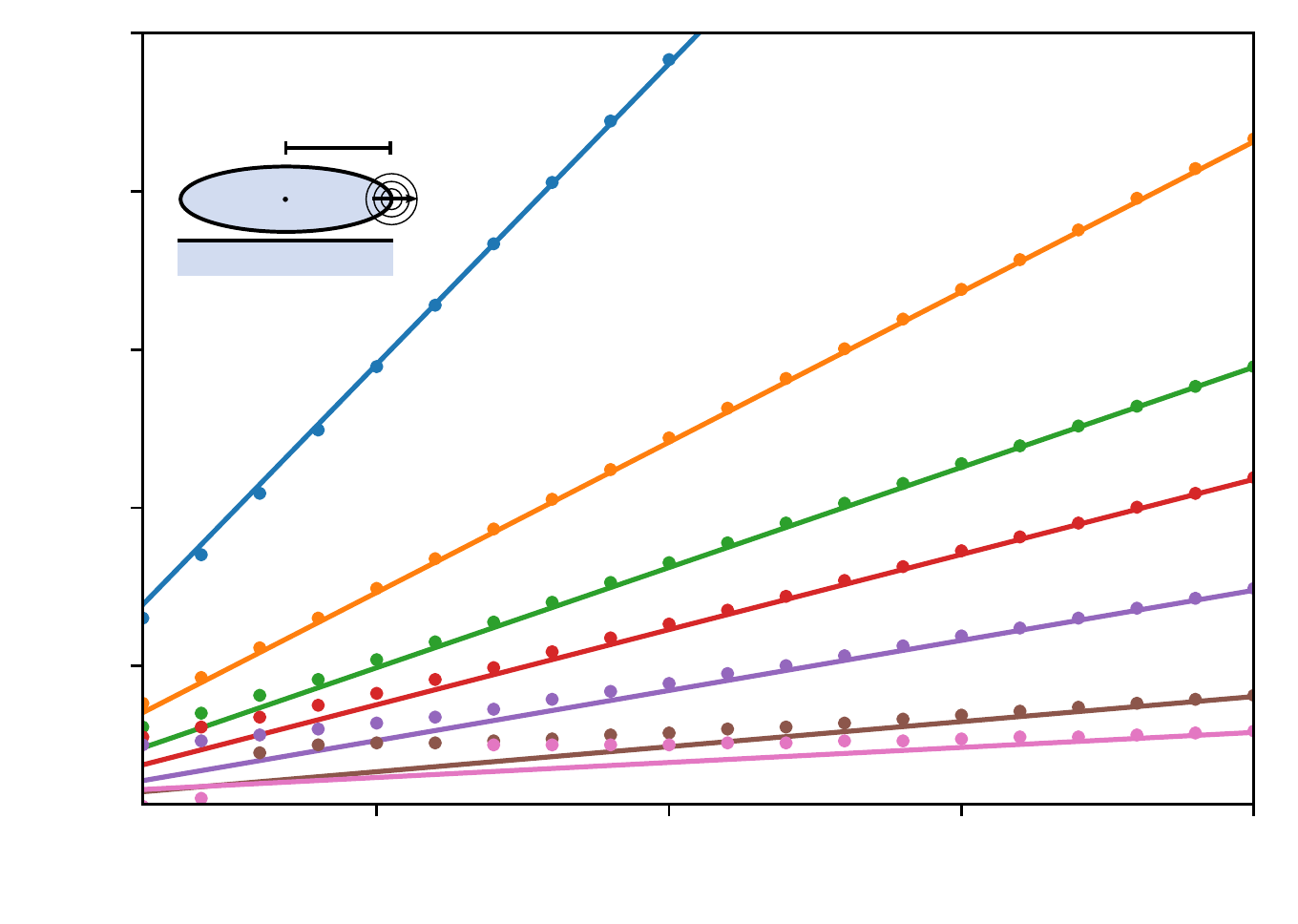}
	\caption{Observed $\kmax$ (according to \eqref{eq:kmax of lmax}, points) and the fit (according to \eqref{eq:kmax rx}, lines) for $\fatr_d\parallel\unite_x\parallel\fatd$. Each line refers to a fixed separation between point current source and particle center $|k\fatr_d|$.}
	\label{fig:kmax vs lmax}
\end{figure}

The goal is to investigate how the convergence of the plane wave spectrum depends on the size of the particle. In fact, the plane wave spectrum and thereby its convergence depends on the actual induced current distribution $\fatj(\fatr)$ in the particle, which we do not know. However, as the convergence of $g^-_{\lmax, j}(\alpha, \kappa)$ is limited by those induced current source elements $\fatj(\fatr)\,\mathrm{d}^3\fatr$ that are located far away from the particle center, a conservative estimate of $\kmax$ can be obtained by modelling the current distribution $\fatj(\fatr)$ with a single point dipole current source located on the surface of the circumscribing sphere \cite{Cappellin_RS_2008}:

\begin{equation}
\mathbf{j}(\fatr) = \delta^3(\fatr-\fatr_d) \, \fatd,
\end{equation}
where the unit vector $\fatd$ refers to the dipole orientation and $\left|\fatr_d\right|=R$, compare figure~\ref{fig:single dipole}.
We will next check how the separation of the infinitesimal current source from the particle center affects the convergence rate.
Figure \ref{fig:g comparison} shows a graphical representation of the plane wave spectrum for a dipole source with $k\fatr_d=(1,0,0)$ and $\fatd=(1,1,0)/\sqrt{2}$. One can clearly see that in the image center, where $\kappa$ is small, the plane wave spectra approach the exact spectrum faster with growing  $\lmax$ compared to the off-centered regions where $\kappa$ is large. This has also been observed by Cappellin et al. \cite{Cappellin_RS_2008}.
Figure \ref{fig:quadratic residual} shows the azimuthally averaged quadratic error as a function of $\kappa$. Indeed, the quadratic residual grows with $\kappa$ and the regime where $\MSE<1$ grows with $\lmax$. 
This increase of $\kmax$ with $\lmax$  is shown in figure \ref{fig:kmax vs lmax} for various $\fatr_d$.

For a fixed dipole location and orientation, $\kmax$ as a function of $\lmax$ can be fitted by a straight line. In principle, the slope and the $y$-intercept of this line depend on the dipole position and orientation. As expected, $\kmax$ grows when the current source moves closer to the particle center, because here the plane wave spectrum $g^-_{\lmax,j}$ converges faster to the exact spectrum. 
 
Regarding the direction of $\fatr_d$, we focus on the case of a lateral displacement, $\fatr_d\perp\unite_z$ -- for the following reason: The truncation of the Sommerfeld integral is only critical for the overall accuracy for flat particles close to the substrate. For non-flat particles, the separation of the particle center to the planar interface leads to a suppression of contributions with large $\kappa$, which correspond to evanescent waves decaying fast with growing $z$-distance to the particle center. But for flat particles, the actual limit of $\kmax$ is defined by the lateral extent of the particle as opposed to the vertical extent.

In the case of a horizontal displacement of the current source, $\kmax$ is always larger for $\fatd\perp\fatr_d$ (not shown) compared to $\fatd\parallel\fatr_d$ (figure \ref{fig:kmax vs lmax}). The critical case is thus $\fatr_d \perp \hat{\mathbf{e}}_z$ with $\fatd\parallel\fatr_d$. We fit the observed behavior of $\kmax$ as a function of $\lmax$ with a phenomenological formula. In the range of $0.5\lesssim|k\fatr_d|\lesssim 10$ and $\lmax \lesssim 20$,
\begin{equation}
\label{eq:kmax rx}
\kmax = (0.38 \lmax + 1)R^{-1} + 0.03 k^2R
\end{equation}
with $R=|\fatr_d|$ provides a reasonable fit, compare figure \ref{fig:kmax vs lmax}.

\begin{figure}[tp]
	\centering
	\makebox[\textwidth][c]
	{\includegraphics[width=140 mm]{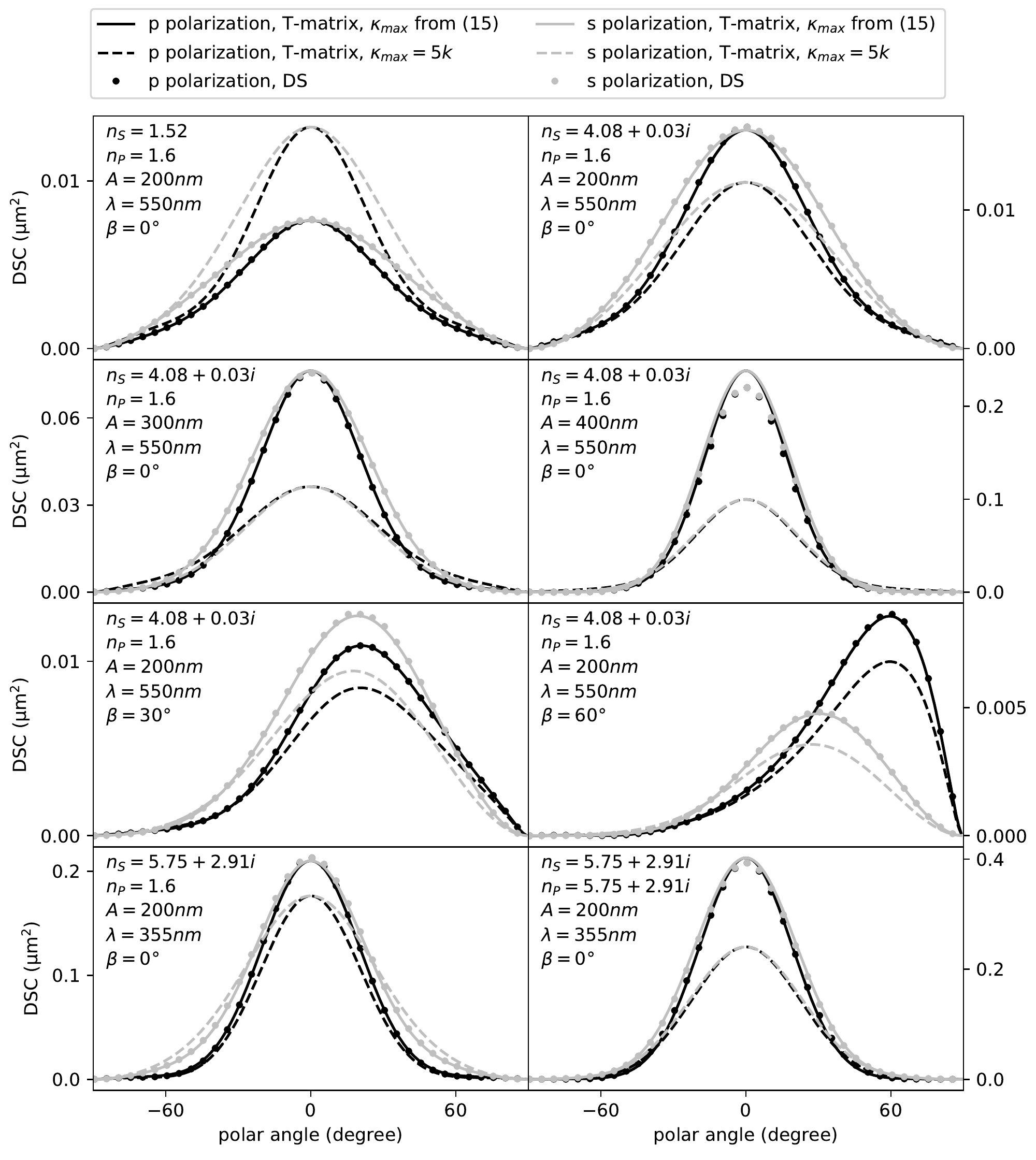}}
    
	\caption{Differential scattering cross section of an oblate spheroid on a plane substrate into the $xz$-plane. The excitation is given by a plane wave polarized into the $y$-direction (``TE'') or into the $x$-direction (``TM''). The spheroid's short semi axis is fixed to $C=50\,\mathrm{nm}$ whereas the size of the long semi axes $A=B$ is indicated near the respective curves, together with the substrate ($n_S$) and particle ($n_P$) refractive index and the plane wave's vacuum wavelength $\lambda$ and polar angle of incidence $\beta$. In the $T$-matrix calculations, $\lmax=10$ was used. The solid lines were computed using truncated Sommerfeld integrals according to \eqref{eq:kmax rx}, whereas the dashed lines refer to a truncation at $\kmax=5k$ which is larger. The dots refer to reference calculations using the discrete sources (DS) method.}
	\label{fig:tmat vs ds}
\end{figure}

\begin{figure}[t!]
	\centering
	\includegraphics[width=0.95\textwidth]{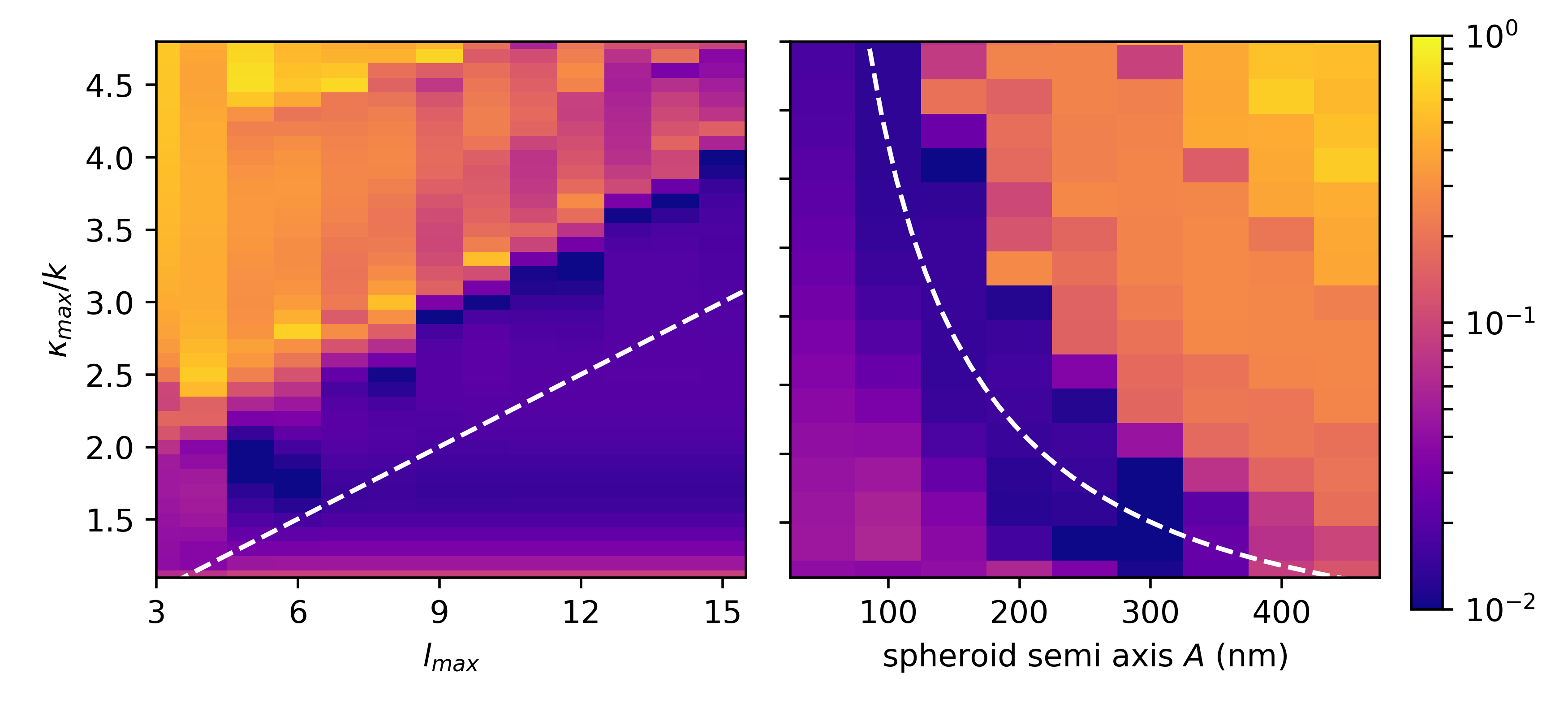}
	\caption{Relative deviation between $T$-matrix and reference simulations for an oblate spheroid ($n_P=1.6$) on a substrate ($n_S=4.08+0.03\mathrm{i}$) under normal incidence at $\lambda=550\,\mathrm{nm}$. In the left panel, the long spheroid semi axis is fixed to $A=200\,\mathrm{nm}$. In the right panel, the truncation multipole degree is fixed to $\lmax=10$. The white dashed lines indicate the estimate of $\kmax$ according to \eqref{eq:kmax rx}.}	\label{fig:tmatrixvsdserror}
\end{figure}

\section{Application examples}
\label{sec:application examples}
In order to explore the application of \eqref{eq:kmax rx} to a real scattering problem, we studied light scattering by an oblate spheroid on a substrate. The axis of revolution is given by the $z$-axis, which is normal to the substrate surface. We denote the spheroid's half axis in $z$-direction by $C$ (fixed to $50\nm$), the half axis in the transverse directions by $A$, the particle's refractive index by $n_P$, the substrate's refractive index by $n_S$ and the incident plane wave's vacuum wavelength by $\lambda$ and its angle of incidence by $\beta$.
In order to compute the scattered far field including the particle-substrate interaction we used SMUTHI, a new Python package for the simulation of scattering particles in layered media, which is available for free download \cite{smuthi2017}. For the simulation of the spheroid's $T$-matrix, the NFM-DS Fortran code based on the null-field method with discrete sources was used \cite{Doicu1997, DoicuA2006}. 
The maximal multipole degree was set to $\lmax=10$, whereas the Sommerfeld integrals were truncated at $\kmax$ according to \eqref{eq:kmax rx}, and for comparison also at a higher value of $\kmax=5k$.
The resulting far fields are then compared to accurate baseline results computed with the discrete sources method \cite{Eremin1999}.

Figure \ref{fig:tmat vs ds} shows the calculated differential scattering cross section for eight different example configurations. The chosen refractive indices correspond to the case of a typical polymer particle or a silicon particle on a glass substrate or on a silicon substrate in the visible or UV region, where silicon is metallic. The parameters of each simulation are shown inside the respective plot panels. In most cases, the agreement between the $T$-matrix based simulations using a Sommerfeld integral truncation according to \eqref{eq:kmax rx} and the DSM reference simulations is very good, except for the case of $A=400\nm$ with an aspect ratio of $C/A=1/8$. In any case, the agreement is significantly better with a correct Sommerfeld integral truncation compared to the case of a truncation at a too large wavenumber $\kmax=5k$.

For the case of $n_P=1.6$, $n_S=4.08+0.03\ui$, $\lambda=550\nm$ and $\beta=0^\circ$, we quantify the relative deviation of the $T$-matrix based simulations from the reference simulations with respect to the $L^2$-norm of the differential scattering cross section. The resulting relative error, as a function of $\lmax$ and $\kmax$ is shown in figure \ref{fig:tmatrixvsdserror}. One can clearly see that for a fixed $A$ and $\lmax$, the error increases significantly for $\kmax$ larger than some critical value, and that this critical value grows linearly with $\lmax$ (see left panel). The estimate \eqref{eq:kmax rx} for  $\kmax$ is shown as white dashed lines. Further, for fixed $\lmax$ the critical $\kmax$ decreases rapidly with increasing lateral semi axis $A$ (see right panel).
In all cases, the estimated $\kmax$ is well below the critical $\kmax$, and can thus be used for valid simulations.

\section{Discussion}
With \eqref{eq:kmax rx} we have suggested a formula for a conservative estimate of $\kmax$, that is the in-plane wavenumber at which the Sommerfeld integral needs to be truncated in order to avoid contributions from the diverging spherical wave expansion in the near field zone. For that purpose, we modelled the generic induced current distribution inside the scattering particle with a single point dipole located at the outermost part of the particle volume, i.e., the position with the maximal distance to the particle center. 

One consequence of this approach is that the estimated $\kmax$ can be smaller than the actual optimal truncation, as the overall scattering response of the particle is not only determined by the outermost volume elements, but also from those located more towards the particle center, which correspond to a smaller $|\fatr_d|$ and thereby to a larger $\kmax$. In this sense, \eqref{eq:kmax rx} is a conservative estimate. This can also be seen in figure \ref{fig:tmatrixvsdserror}, where the estimated $\kmax$ is well below the critical $\kmax$.

It is important to note that a truncation of the Sommerfeld integral according to \eqref{eq:kmax rx} is neither sufficient nor necessary for an overall accurate simulation result. The challenge is rather to pick $\lmax$ and $\kmax$ such that each of the following three requirements is satisfied:
\begin{enumerate}
	\item The truncation multipole degree $\lmax$ is large enough such that the spherical wave expansion of the conventional $T$-matrix method converges to the desired accuracy. This can be assured for example by the Wiscombe criterion or some refined prescription of similar kind \cite{Wiscombe1980,Neves_OL_2012}.
	\item The truncation in-plane wavenumber $\kmax$ is large enough such that the relevant part of the evanescent wave spectrum is included in the treatment. This depends on the distance $\Delta z$ between particle center and substrate, as the contribution of the evanescent waves is damped with $\sim \exp(2\mathrm{i} \Delta z k_z)$ where the imaginary part of $k_z$ grows with $\kappa>k$. If this criterion leads to a $\kmax$ smaller than suggested by \eqref{eq:kmax rx}, there is no need to use the larger value.
	\item The truncation wavenumber $\kmax$ is not much larger than the domain where the plane wave spectrum has converged. This can be assured by using \eqref{eq:kmax rx}.
\end{enumerate}

Finally, we note that the here presented analysis does not only hold for particles near an interface, but in general for the near field reconstruction by means of a transformation from a spherical wave expansion to a plane wave expansion, for example in the context of multiple scattering by non spherical particles with overlapping circumscribing spheres \cite{Theobald2017}.

\section{Conclusions}
We have presented a formula for the estimation of a suitable truncation scale $\kmax$ of Sommerfeld integrals appearing in the simulation of light scattering by flat structures on a substrate with the $T$-matrix method. By comparing the calculated differential scattering cross section to accurate baseline results, we have confirmed that the estimate is valid. We hope that this work will turn out useful in propagating the  $T$-matrix method to a broader range of applications, where a careful treatment of the scattered near field interaction between scattering particles and planar interfaces is essential.

\section{Acknowledgements}
AE and DT acknowledge support from the Karlsruhe School of Optics \& Photonics (KSOP). GG gratefully acknowledges support from the Helmholtz Postdoc Program. This work was funded by the DFG through the priority programme 1839 ``Tailored disorder''. 

\bibliographystyle{model1-num-names}
\bibliography{si_truncation}

\end{document}